\documentclass[conference,a4paper,twocolumn]{IEEEtran}

\IEEEoverridecommandlockouts

\usepackage{cite}

\ifCLASSINFOpdf
\else
\fi
\usepackage[cmex10]{amsmath}

\usepackage{tipa}
\usepackage[pdftex]{graphicx}
\usepackage{epsfig}
\usepackage{epstopdf}
\usepackage[flushleft]{threeparttable}
\usepackage{flushend}

\begin{document}

\title{Identification of Hypokinetic Dysarthria Using Acoustic Analysis of Poem Recitation}

\author{\IEEEauthorblockN{Jan~Mucha\IEEEauthorrefmark{1},
													Zoltan~Galaz\IEEEauthorrefmark{1},
													Jiri~Mekyska\IEEEauthorrefmark{1},
													Tomas~Kiska\IEEEauthorrefmark{1},
													Vojtech~Zvoncak\IEEEauthorrefmark{1},
													Zdenek~Smekal\IEEEauthorrefmark{1}\\
													Ilona~Eliasova\IEEEauthorrefmark{2}\IEEEauthorrefmark{3},
													Martina~Mrackova\IEEEauthorrefmark{2}\IEEEauthorrefmark{3},
													Milena~Kostalova\IEEEauthorrefmark{3},
													Irena~Rektorova\IEEEauthorrefmark{2}\IEEEauthorrefmark{3}\\ 
													Marcos~Faundez-Zanuy\IEEEauthorrefmark{4}, and 
													Jesus~B.~Alonso-Hernandez\IEEEauthorrefmark{5}}
				\IEEEauthorblockA{\IEEEauthorrefmark{1}Department of Telecommunications and SIX Research Centre, Brno University of Technology\\Technicka~10, 61600 Brno, Czech Republic}
				\IEEEauthorblockA{\IEEEauthorrefmark{2}First Department of Neurology, Masaryk University and St. Anne's University Hospital\\Pekarska 53, 65691 Brno, Czech Republic}
				\IEEEauthorblockA{\IEEEauthorrefmark{3}Applied Neuroscience Research Group, Central European Institute of Technology\\Masaryk University, Kamenice 5, 62500 Brno, Czech Republic}
				\IEEEauthorblockA{\IEEEauthorrefmark{4}Escola Superior Politecnica, Tecnocampus\\Avda. Ernest Lluch 32, 08302 Mataro, Barcelona, Spain}
				\IEEEauthorblockA{\IEEEauthorrefmark{5}Institute for Technological Development and Innovation in Communications (IDeTIC)\\University of Las Palmas de Gran Canaria, 35001~Las Palmas de Gran Canaria, Spain}
				\thanks{This work was supported by the grant of the Czech Ministry of Health 16-30805A (Effects of non-invasive brain stimulation on hypokinetic dysarthria, micrographia, and brain plasticity in patients with Parkinson’s disease) and the following projects: SIX (CZ.1.05/2.1.00/03.0072), LOl401, FEDER and MICINN, TEC2016-77791-C4-2-R. For the research, infrastructure of the SIX Center was used.}}
				
\maketitle

\begin{abstract}
Up to 90\,\% of patients with Parkinson's disease (PD) suffer from hypokinetic dysarthria (HD). In this work, we analysed the power of conventional speech features quantifying imprecise articulation, dysprosody, speech dysfluency and speech quality deterioration extracted from a~specialized poem recitation task to discriminate dysarthric and healthy speech. For this purpose, 152 speakers (53 healthy speakers, 99 PD patients) were examined. Only mildly strong correlation between speech features and clinical status of the speakers was observed. In the case of univariate classification analysis, sensitivity of 62.63\,\% (imprecise articulation), 61.62\,\% (dysprosody), 71.72\,\% (speech dysfluency) and 59.60\,\% (speech quality deterioration) was achieved. Multivariate classification analysis improved the classification performance. Sensitivity of 83.42\,\% using only two features describing imprecise articulation and speech quality deterioration in HD was achieved. We showed the promising potential of the selected speech features and especially the use of poem recitation task to quantify and identify HD in PD.
\end{abstract}


\IEEEpeerreviewmaketitle

\section{Introduction}
\label{S:1}

Parkinson's disease (PD) is a~frequent neurodegenerative disorder~\cite{Rijk2000}. Prevalence rate of PD was estimated to approximately 1.5\,\% (people aged over 65 years), whereas the risk of PD onset increases with age~\cite{Sapir2008}. The exact pathophysiological cause of PD is still yet to be found, however the most significant biological finding associated with the disease is a~rapid degeneration of dopaminergic cells in \textit{substancia nigra pars compacta}\cite{Hornykiewicz1998}. The cardinal motor symptoms of PD comprise tremor in rest, progresive bradykinesia, muscular rigidity/stiffness~\cite{Mekyska2011b_eng}, etc. Patients with PD often develop several non-motor symptoms~\cite{Brabenec2017} such as sleep disorders, cognitive deficits, depression, dementia, etc. 

According to~\cite{Ho1999a}, up to 90\,\% of PD patients suffer from motor speech disorder named hypokinetic dysarthria (HD)~\cite{Darley1969}. HD affects different subsystems involved in the formation of speech~\cite{Ramig2008, Hartelius1994} such as articulation, phonation, prosody and speech fluency~\cite{Mekyska2011b_eng}. Increased acoustic noise, reduced voice intensity, harsh breathy voice quality, increased voice nasality, reduced variability of pitch and loudness, speech rate abnormalities, imprecise articulation, unintentional introduction of pauses, rapid repetition of words or syllables and sudden deceleration/acceleration in speech have been observed in patients with PD~\cite{Skodda2009, Mekyska2015, Galaz2016, Brabenec2017}.

For quantification and identification of HD in PD, a~wide range of speech tasks such as sustained vowel phonation~\cite{Mekyska2015a, Tsanas2010}, fast syllable repetition tasks~\cite{Skodda2011, Rusz2013}, read text~\cite{Skodda2011c, Eliasova2013, Galaz2016, Elfmarkova2016}, and running speech~\cite{Skodda2009, Skodda2011c, Rusz2013b}, etc. has been employed. Additionally, HD has been analysed using conventional, clinically interpretable speech processing techniques based on the description of: a) articulation (reduced mobility of the articulatory organs)~\cite{Sapir2010, Skodda2011, Rusz2011, Smekal2015a}, b) speech prosody (reduced variability of pitch and loudness)~\cite{Skodda2009, Skodda2011c, Mekyska2015, Galaz2016}, c) speech fluency (speech rate and pausing abnormalities)~\cite{Skodda2009, Skodda2011c, Rusz2011, Galaz2016}, d) speech quality (increased level of voice tremor)~\cite{Tsanas2010, Tsanas2010b, Mekyska2015a, Smekal2015a}. Further information can be found in our recent article~\cite{Brabenec2017}. 

To our best knowledge, despite our previous work~\cite{Galaz2016}, quantification and identification of HD in PD using acoustic analysis of a~poem recitation has not been performed. Therefore, the aim of this work is to follow and extend our previous research and evaluate a~potential of conventional speech features describing imprecise articulation, dysprosody, speech fluency deficits and speech quality deterioration in HD extracted from a~specialized poem recitation task, which requires additional rhythmical patterns, stress and intonation, to discriminate between healthy and dysarthric speech in patients with idiopathic PD.

\section{Materials and methods}
\label{S:2}

\subsection{Dataset}
\label{S:2_1}

For the purpose of this study, 152 Czech native speakers were examined. The group of speakers consisted of 53 healthy controls (HC) with (mean~$\pm$~std) age: 63.87~$\pm$~9.28 years and 99 PD patients with (mean~$\pm$~std) age: 67.50~$\pm$~8.08 years, PD duration: 7.47~$\pm$~4.17 years, UPDRS~III (Unified Parkinson's disease rating scale, part III: evaluation of motor function)~\cite{Fahn1987}: 23.58~$\pm$~12.16 and LED (L-dopa equivalent daily dose)~\cite{Lee2010}: 1002.74~$\pm$~557.25 mg. All study participants were enrolled at the First Department of Neurology, St. Anne's University Hospital in Brno, Czech Republic. The PD patients were examined approximately 1~hour after their regular dopaminergic medication (L-dopa dose~\cite{Lee2010}). All PD patients signed an informed consent form approved by the local ethics committee.

\subsection{Data acquisition}
\label{S:2_2}

During the speech signals acquisition, all speakers performed the specifically-designed poem recitation task~\cite{Galaz2016}: Czech (original) version\,--\,\textit{Chcete vid\v{e}t velk\'{y} lov? Budu lovit v~d\v{z}ungli slov. Osedl\'{a}m si Pegasa, chyt\'{\i}m b\'{a}se\v{n} do lasa!}, English (translation) version\,--\,\textit{Would you like to see a~big hunt? I~will be hunting in a~jungle of words. I~will saddle the Pegasus, I~will catch a~poem into a~lasso}. All speakers were asked to first read and try to recite a~poem for themselves. They recited the poem into a~microphone afterwards.

\subsection{Speech features}
\label{S:2_3}

To robustly and complexly describe HD, We extracted a~variety of conventional speech features~\cite{Skodda2009, Mekyska2015, Galaz2016, Brabenec2017}: a) imprecise articulation\,--\,first three formant frequencies (F1\,--\,F3) and their bandwidths (B1\,--\,B3); b) dysprosody\,--\,fundamental frequency (F0), pitch period entropy (PPE), short-time energy (STE), and Teager energy operator (TEO); c) speech rate and fluency deficits\,--\,number of voice breaks (NVB), degree of voice breaks (DVB), total pause time (TPT), total pause time (pauses longer than 50 ms) (TPT\,50), articulation rate (AR), and speech index of rhythmicity (SPIR); and d) speech quality deterioration\,--\,harmonics-to-noise ratio (HNR), median of power spectrum density (MPSD), spectrum flux (SF), zero-crossing rate (ZCR), and voice turbolence index (VTI). For further information, see~\cite{Mekyska2015}.

Next, we described the statistical properties of the features using: range (R), interpercentile range (IPR), interdecile range (IDR), interquartile range (IQR), mean, median, std, skewness, kurtosis, coefficient of variation (CV), $5^{\mathrm{th}}$ percentile ($5^{\mathrm{th}}$ p), $1^{\mathrm{st}}$ quartile ($1^{\mathrm{st}}$ q), $3^{\mathrm{rd}}$ quartile ($3^{\mathrm{rd}}$ q), $95^{\mathrm{th}}$ percentile ($95^{\mathrm{th}}$ p) and slope of linear regression (sLR)~\cite{Mekyska2015}.

\subsection{Statistical analysis}
\label{S:2_4}

After the feature extraction step, we investigated the strength of a~linear/monotonic relationship between the speech features and clinical status of the speakers using Pearson's correlation coefficient ($r_p$) and Spearman's correlation coefficient ($r_s$). The significance level of correlation $\alpha\,=\,0.05$ was selected. Next, to evaluate individual discrimination power of the features, univariate binary classification (PD/HC) models (stratified 10-fold validation with 20 repetitions) based on random forests (RF)~\cite{Breiman2001} classifier were designed.

Consequently, to propose the classification models with the maximum discrimination power (HD is a~multimodal speech disorder), multivariate binary classification (PD/HC) models (stratified 10-fold validation with 20 repetitions) using RF classifier~\cite{Breiman2001} were designed. To select the most suitable combination of the features~\cite{Guyon2006}, sequential floating forward selection (SFFS) algorithm~\cite{Pohjalainen2014} was used.

RF classifier was selected considering its ability to deal with high-dimensional and highly correlated data with complex interactions, which is often the case in this field of science. The training and testing features were normalized before classification on a~per-feature basis ($\mu=0$, $\sigma=1$). The classification performance was evaluated by Matthew's correlation coefficient~\cite{Matthews1975, Jurman2012}, classification accuracy (ACC), sensitivity (SEN) and specificity (SPE)~\cite{Altman1994, Galaz2016}.

\section{Results}
\label{S:3}

\begin{figure*}[htb!]
  \centering
  \includegraphics[width=0.9\textwidth]{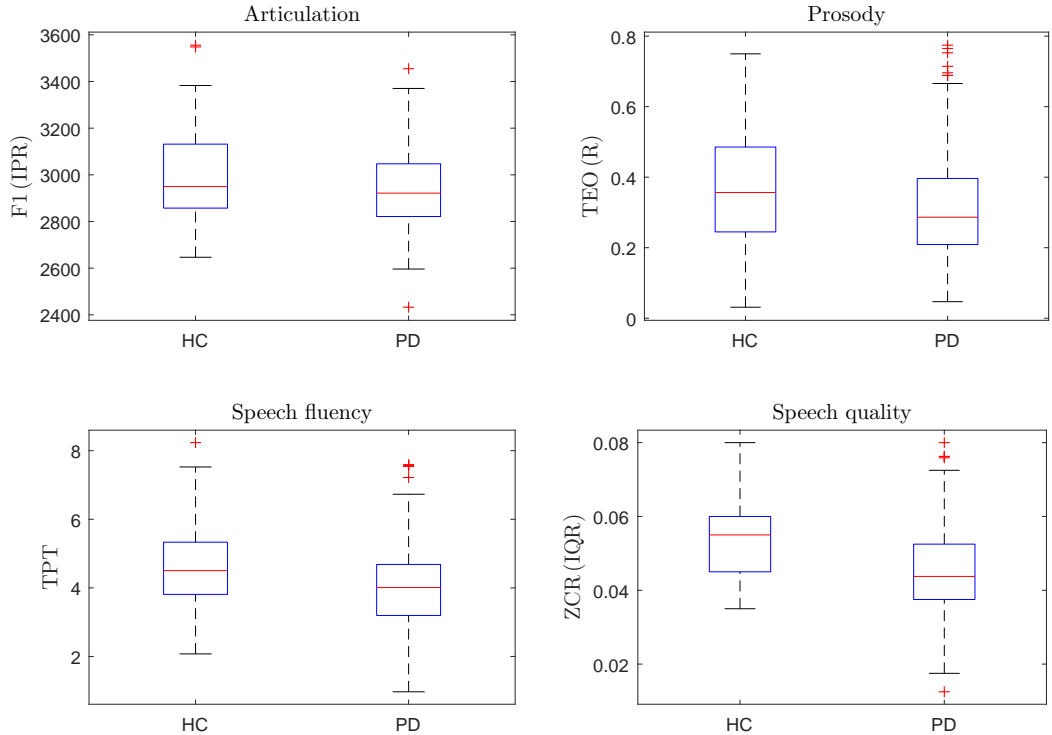}
  \caption{Box plots of the features with the highest MCC for each analysed speech dimension (articulation, prosody, speech fluency and speech quality) for HC and patients with PD. Speech features notation: F1\,(IPR)~--~interpercentile range of the first formant frequency; TEO\,(R)~--~range of Teager energy operator; TPT~--~total pause time; ZCR\,(IQR)~--~interquartile range of zero-crossing rate; HC~--~healthy controls; PD~--~patients with idiopathic Parkinson's disease.}
  \label{fig:Box_plots}
\end{figure*}

The results of the univariate/multivariate analysis are summarized in Table~\ref{tab:RES_table} (the table shows the features with the highest MCC for each speech dimension, see Section~\ref{S:2_3}). Regarding the correlation analysis, only mildly strong correlation between speech features and clinical status of the speakers (PD/HC) was observed. Regarding the univariate classification analysis, the following results were achieved (SEN): a) features based on the quantification of imprecise articulation\,--\,$\mbox{F1\,(IPR)} = 62.63\,\%$; b) features based on quantification of speech prosody impairment\,--\,$\mbox{TEO\,(R)} = 61.62\,\%$; c) features based on quantification of speech rate, regularity and fluency deficits\,--\,$\mbox{TPT} = 71.72\,\%$; and finally d) features based on quantification of voice and speech quality deterioration\,--\,$\mbox{ZCR\,(IQR)} = 59.60\,\%$. Box plots of the features with the highest MCC (univariate models) for each speech dimension are visualized in Figure~\ref{fig:Box_plots}. With respect to the multivariate classification analysis, the SEN of $83.42\,\%$ using the combination of MPDS\,(median), F1\,(IDR) was achieved.

\begin{table*}[htb!]
  \centering
  \begin{threeparttable}
    \caption{Results of the correlation and univariate/multivariate classification analysis.}
    \label{tab:RES_table}
    \footnotesize
    \centering
    
    \begin{tabular}{l l c c c c c c c c}
			\hline
			\hline
			\multicolumn{10}{c}{Univariate analysis} \\
			speech dim. & speech feature & $r_p$ & $p_p$ & $r_s$ & $p_s$ & MCC & ACC\,[\%] & SEN\,[\%] & SPE\,[\%] \\
			\hline
			
				A & F1\,(IPR) & -0.1428 & 0.0791 & -0.1288 & 0.1136 & 0.2558 & 63.16 & 62.63 & 64.15 \\
				
				P & TEO\,(R) & -0.0886 & 0.2776 & -0.1049 & 0.1982 & 0.2459 & 62.50 & 61.62 & 64.15 \\
				
				S & TPT & -0.1141 & 0.1616 & -0.1757 & 0.0303 & 0.3129 & 67.76 & 71.72 & 60.38 \\
				
				Q & ZCR\,(IQR) & -0.2158 & 0.0076 & -0.3027 & 0.0002 & 0.3347 & 65.13 & 59.60 & 75.47 \\
			
			\hline
			\multicolumn{10}{c}{Multivariate analysis} \\
			speech dim. & speech features & MCC & ACC\,[\%] & SEN\,[\%] & SPE\,[\%] \\
			\hline
			
				Q, A & MPDS\,(median), F1\,(IDR) & 0.5699 & 79.72 & 83.42 & 74.02 \\
			
			\hline
			\hline
    \end{tabular}
    
    \begin{tablenotes}
      \scriptsize
      \item[1] speech dim.~--~speech dimension impaired by the presence of hypokinetic dysarthria; $r_p$~--~Pearson's correlation coefficient; $r_s$~--~Spearman's correlation coefficient; $p_p$~--~significance level of correlation ($r_p$); $p_s$~--~significance level of correlation ($r_s$); MCC~--~Matthew's correlation coefficient; ACC~--~accuracy; SEN~--~sensitivity; SPE~--~specificity; A~--~articulation; P~--~prosody; S~--~speech fluency; Q~--~speech quality; F1\,(IPR)~--~interpercentile range of the first formant frequency; TEO\,(R)~--~range of Teager energy operator; TPT~--~total pause time; ZCR\,(IQR)~--~interquartile range of zero-crossing rate.
    \end{tablenotes}
    
  \end{threeparttable}
\end{table*}

\section{Discussion}
\label{S:4}

Results of this work confirm our previous findings~\cite{Galaz2016} of a~promising potential of the poem recitation task to describe HD in PD. Here, we showed that in addition to prosodic impairment, the poem recitation task is capable of sufficient quantification of reduced mobility of the articulatory organs, speech dysfluency and also speech quality deterioration present in HD. Therefore an inclusion of the poem recitation task in the standardized speech examination protocol is likely to provide additional clinical information about speech disorders associated with HD in patients with PD.

Based on the results summarized in Figure~\ref{fig:Box_plots}, we confirmed the previous findings of impaired mobility of the articulatory organs (imprecise articulation of consonants), reduced variability of speech intensity (monoloudness), speech rate deficits (speech dysfluency) and speech quality deterioration (increased voice tremor) in patients with PD compared to healthy speakers~\cite{Skodda2008, Rusz2013, Mekyska2015, Galaz2016, Brabenec2017}. However, the results need to be verified by subsequent studies.

Furthermore, important observation regarding the multivariate analysis is the fact that multivariate classification model designed using only two speech features, selected by SFFS algorithm, describing imprecise articulation and speech quality deterioration in HD was able to outperform the results of our previous study~\cite{Galaz2016} investigating prosodic speech features extracted from the same poem recitation task in patients with PD. The next step is the analysis of speech features extracted from a~complex speech protocol consisting of classical speech tasks including the poem recitation task.

\section{Conclusion}
\label{S:5}

With respect to the results presented in this work, we conclude that the poem recitation task is capable of precise and robust quantification and identification of HD in PD using a~reasonable number of clinically relevant speech features commonly used in this field of science~\cite{Skodda2008, Skodda2011, Mekyska2015, Brabenec2017}. 

Nevertheless, to generalize these results another study comparing the classification performance of these features using standardized reading tasks and monologue should be performed. Furthermore, the results presented in this paper are based on the analysis of conventional speech features. Investigation of novel, non-conventional features is likely to improve the discrimination power of the classification models. Moreover, gender-differentiated analysis should be considered to emphasize gender-specific aspects of HD in PD.

In our future studies, we aim to perform a~complex analysis of speech tasks including a~poem recitation task and novel, non-conventional speech features to design classification models capable of precise, reliable quantification and identification of HD in PD. Furthermore, we aim to use the same methodology to design regression models capable of objective PD severity estimation, which could provide clinicians with additional power to diagnose, rate and monitor PD.

\bibliographystyle{IEEEtran}
\bibliography{bib_database}

\begin{thebibliography}{10}
\providecommand{\url}[1]{#1}
\csname url@samestyle\endcsname
\providecommand{\newblock}{\relax}
\providecommand{\bibinfo}[2]{#2}
\providecommand{\BIBentrySTDinterwordspacing}{\spaceskip=0pt\relax}
\providecommand{\BIBentryALTinterwordstretchfactor}{4}
\providecommand{\BIBentryALTinterwordspacing}{\spaceskip=\fontdimen2\font plus
\BIBentryALTinterwordstretchfactor\fontdimen3\font minus
  \fontdimen4\font\relax}
\providecommand{\BIBforeignlanguage}[2]{{%
\expandafter\ifx\csname l@#1\endcsname\relax
\typeout{** WARNING: IEEEtran.bst: No hyphenation pattern has been}%
\typeout{** loaded for the language `#1'. Using the pattern for}%
\typeout{** the default language instead.}%
\else
\language=\csname l@#1\endcsname
\fi
#2}}
\providecommand{\BIBdecl}{\relax}
\BIBdecl

\bibitem{Rijk2000}
M.~C. de~Rijk, L.~J. Launer, K.~Berger, M.~M. Breteler, J.~F. Dartigues,
  M.~Baldereschi, L.~Fratiglioni, J.~Lobo, A. Martinez-Lage, C.~Trenkwalder,
  and A.~Hofman, ``Prevalence of {Parkinson'{s}} disease in {Europe}: {A}
  collaborative study of population-based cohorts,'' \emph{Neurology}, vol.~54,
  pp. 21--23, 2000.

\bibitem{Sapir2008}
S.~Sapir, L.~Ramig, and C.~Fox, ``{Speech and swallowing disorders in Parkinson
  disease},'' \emph{Curr. Opin. Otolaryngol. Head Neck Surg.}, vol.~16, no.~3,
  pp. 205--10, 2008.

\bibitem{Hornykiewicz1998}
O.~Hornykiewicz, ``Biochemical aspects of parkinson'{s} disease,''
  \emph{Neurology}, vol.~51, no. 2 Suppl 2, pp. S2--S9, 1998.

\bibitem{Mekyska2011b_eng}
J.~Mekyska, Z.~Smekal, M.~Kostalova, M.~Mrackova, S.~Skutilova, and
  I.~Rektorova, ``Motor aspects of speech impairment in {Parkinson}'{s} disease
  and their assessment,'' \emph{Cesk Slov Neurol N}, vol.~74, no.~6, pp.
  662--668, 2011.

\bibitem{Brabenec2017}
\BIBentryALTinterwordspacing
L.~Brabenec, J.~Mekyska, Z.~Galaz, and I.~Rektorova, ``Speech disorders in
  parkinson's disease: early diagnostics and effects of medication and brain
  stimulation,'' \emph{Journal of Neural Transmission}, pp. 1--32, 2017.
  [Online]. Available: \url{http://dx.doi.org/10.1007/s00702-017-1676-0}
\BIBentrySTDinterwordspacing

\bibitem{Ho1999a}
A.~K. Ho, R.~Iansek, C.~Marigliani, J.~L. Bradshaw, and S.~Gates, ``Speech
  impairment in a large sample of patients with {Parkinson'{s} disease},''
  \emph{J. Behav. Neurol.}, vol.~11, no.~3, pp. 131--137, 1999.

\bibitem{Darley1969}
F.~L. Darley, A.~E. Aronson, and J.~R. Brown, ``{Differential Diagnostic
  Patterns of Dysarthria},'' \emph{J Speech Hear Res}, vol.~12, no.~2, pp.
  246--269, 1969.

\bibitem{Ramig2008}
L.~O. Ramig, C.~Fox, and S.~Sapir, ``Speech treatment for {Parkinson'{s}}
  disease,'' \emph{Expert Rev. Neurother.}, vol.~8, no.~2, pp. 297--309, 2008.

\bibitem{Hartelius1994}
L.~Hartelius and P.~Svensson, ``{Speech and Swallowing Symptoms Associated with
  {Parkinson'{s} disease} and Multiple Sclerosis: a Survey},'' \emph{Folia
  Phoniatr Logop}, vol.~46, no.~1, pp. 9--17, 1994.

\bibitem{Skodda2009}
S.~Skodda, H.~Rinsche, and U.~Schlegel, ``Progression of dysprosody in
  {Parkinson'{s} disease} over time--a longitudinal study,'' \emph{Mov.
  Disord.}, vol.~24, pp. 716--722, 2009.

\bibitem{Mekyska2015}
J.~Mekyska, E.~Janousova, P.~Gomez-Vilda, Z.~Smekal, I.~Rektorova, I.~Eliasova,
  M.~Kostalova, M.~Mrackova, J.~B. Alonso-Hernandez, M.~Faundez-Zanuy, and
  K.~L. de~Ipi{\~n}a, ``Robust and complex approach of pathological speech
  signal analysis,'' \emph{Neurocomputing}, vol. 167, pp. 94--111, 2015.

\bibitem{Galaz2016}
Z.~Galaz, J.~Mekyska, Z.~Mzourek, Z.~Smekal, I.~Rektorova, I.~Eliasova,
  M.~Kostalova, M.~Mrackova, and D.~Berankova, ``Prosodic analysis of neutral,
  stress-modified and rhymed speech in patients with parkinson's disease,''
  \emph{Comput. Methods. Programs. Biomed.}, vol. 127, pp. 301 -- 317, 2016.

\bibitem{Mekyska2015a}
J.~Mekyska, Z.~Galaz, Z.~Mzourek, Z.~Smekal, I.~Rektorova, I.~Eliasova,
  M.~Kostalova, M.~Mrackova, D.~Berankova, M.~Faundez-Zanuy, K.~L. de~Ipiña,
  and J.~B. Alonso-Hernandez, ``Assessing progress of parkinson's disease using
  acoustic analysis of phonation,'' \emph{2015 4th International Work
  Conference on Bioinspired Intelligence (IWOBI)}, pp. 111--118, June 2015.

\bibitem{Tsanas2010}
A.~Tsanas, M.~A. Little, P.~E. McSharry, and L.~O. Ramig, ``Nonlinear speech
  analysis algorithms mapped to a standard metric achieve clinically useful
  quantification of average {Parkinson'{s} disease} symptom severity,''
  \emph{J. R. Soc. Interface}, vol.~8, no.~59, pp. 842--855, 2010.

\bibitem{Skodda2011}
S.~Skodda, W.~Visser, and U.~Schlegel, ``Vowel articulation in {Parkinson'{s}}
  disease,'' \emph{J. Voice}, vol.~25, no.~4, pp. 467--472, 2011.

\bibitem{Rusz2013}
J.~Rusz, R.~Cmejla, T.~Tykalova, H.~Ruzickova, J.~Klempir, V.~Majerova,
  J.~Picmausova, J.~Roth, and E.~Ruzicka, ``Imprecise vowel articulation as a
  potential early marker of {Parkinson'{s} disease}: effect of speaking task,''
  \emph{J. Acoust. Soc. Am.}, vol. 134, no.~3, pp. 2171--2181, 2013.

\bibitem{Skodda2011c}
S.~Skodda, W.~Visser, and U.~Schlegel, ``Gender-related patterns of dysprosody
  in {Parkinson'{s}} disease and correlation between speech variables and motor
  symptoms,'' \emph{J. Voice}, vol.~25, no.~1, pp. 76--82, 2011.

\bibitem{Eliasova2013}
I.~Eliasova, J.~Mekyska, M.~Kostalova, R.~Marecek, Z.~Smekal, and I.~Rektorova,
  ``Acoustic evaluation of short-term effects of repetitive transcranial
  magnetic stimulation on motor aspects of speech in {Parkinson}'{s} disease,''
  \emph{J. Neural Transm.}, vol. 120, no.~4, pp. 597--605, 2013.

\bibitem{Elfmarkova2016}
N.~Elfmarkova, M.~Gajdos, M.~Mrackova, J.~Mekyska, M.~Mikl, and I.~Rektorova,
  ``Impact of parkinson's disease and levodopa on resting state functional
  connectivity related to speech prosody control,'' \emph{Parkinsonism. Relat.
  Disord.}, vol. 22 Suppl 1, pp. S52--5, 2016.

\bibitem{Rusz2013b}
J.~Rusz, R.~Cmejla, H.~Ruzickova, J.~Klempir, V.~Majerova, J.~Picmausova,
  J.~Roth, and E.~Ruzicka, ``Evaluation of speech impairment in early stages of
  {Parkinson'{s}} disease: a prospective study with the role of
  pharmacotherapy,'' \emph{J. Neural Transm.}, vol. 120, no.~2, pp. 319--329,
  2013.

\bibitem{Sapir2010}
S.~Sapir, L.~O. Ramig, J.~L. Spielman, and C.~Fox, ``Formant centralization
  ratio ({FCR}): a proposal for a new acoustic measure of dysarthric speech,''
  \emph{J. Speech Lang. Hear. Res.}, vol.~53, no.~1, pp. 1--20, 2010.

\bibitem{Rusz2011}
J.~Rusz, R.~Cmejla, H.~Ruzickova, and E.~Ruzicka, ``Quantitative acoustic
  measurements for characterization of speech and voice disorders in early
  untreated {Parkinson'{s}} disease,'' \emph{The Journal of the Acoustical
  Society of America}, vol. 129, no.~1, pp. 350--367, 2011.

\bibitem{Smekal2015a}
Z.~Smekal, J.~Mekyska, Z.~Galaz, Z.~Mzourek, I.~Rektorova, and
  M.~Faundez{-}Zanuy, ``Analysis of phonation in patients with {Parkinson}'{s}
  disease using empirical mode decomposition,'' in \emph{2015 International
  Symposium on Signals, Circuits and Systems (ISSCS)}, July 2015, pp. 1--4.

\bibitem{Tsanas2010b}
A.~Tsanas, M.~Little, P.~McSharry, and L.~Ramig, ``Accurate telemonitoring of
  {Parkinson'{s}} {Disease} progression by noninvasive speech tests,''
  \emph{IEEE Trans. Bio-Med. Eng.}, vol.~57, no.~4, pp. 884--893, 2010.

\bibitem{Fahn1987}
S.~Fahn and R.~L. Elton, \emph{UPDRS Development Committee (1987) Unified
  {Parkinson'{s}} Disease Rating Scale}.\hskip 1em plus 0.5em minus 0.4em\relax
  Recent developments in {Parkinson'{s}} Disease. Macmillan, Florham Park,
  1987.

\bibitem{Lee2010}
J.~Y. Lee, J.~W. Kim, W.~Y. Lee, J.~M. Kim, T.~B. Ahn, H.~J. Kim, J.~Cho, and
  B.~S. Jeon, ``Daily dose of dopaminergic medications in {Parkinson'{s}
  disease}: clinical correlates and a posteriori equation,'' \emph{Neurol.
  Asia}, vol.~15, no.~2, pp. 137--143, 2010.

\bibitem{Breiman2001}
L.~Breiman, ``Random forests,'' \emph{Mach. Learn.}, vol.~45, no.~1, pp. 5--32,
  2001.

\bibitem{Guyon2006}
I.~Guyon, S.~Gunn, M.~Nikravesh, L.~A. Zadeh, and Eds, \emph{{Feature
  Extraction: Foundations and Applications}}.\hskip 1em plus 0.5em minus
  0.4em\relax New York: Springer, 2006.

\bibitem{Pohjalainen2014}
J.~Pohjalainen, O.~Rasanen, and S.~Kadioglu, ``Feature selection methods and
  their combinations in high-dimensional classification of speaker likability,
  intelligibility and personality traits,'' \emph{Comput. Speech Lang.}, 2014.

\bibitem{Matthews1975}
B.~W. Matthews, ``Comparison of the predicted and observed secondary structure
  of {T}4 phage lysozyme,'' \emph{Biochim. Biophys. Acta (BBA)}, vol. 405,
  no.~2, pp. 442--51, 1975.

\bibitem{Jurman2012}
G.~Jurman, S.~Riccadonna, and C.~Furlanello, ``A comparison of {MCC} and {CEN}
  error measures in multi-class prediction,'' \emph{{PLoS} {ONE}}, vol.~7,
  no.~8, p. e41882, aug 2012.

\bibitem{Altman1994}
\BIBentryALTinterwordspacing
D.~G. Altman and J.~M. Bland, ``Statistics notes: Diagnostic tests 1:
  sensitivity and specificity,'' \emph{BMJ}, vol. 308, no. 6943, p. 1552, 1994.
  [Online]. Available: \url{http://www.bmj.com/content/308/6943/1552}
\BIBentrySTDinterwordspacing

\bibitem{Skodda2008}
S.~Skodda and U.~Schlegel, ``Speech rate and rhythm in parkinson'{s} disease,''
  \emph{Mov. Disord.}, vol.~23, no.~7, pp. 985--92, 2008.

\end{thebibliography}

\end{document}